\documentclass[11pt]{article}
\usepackage{moriond,epsfig}

\def\Journal#1#2#3#4{{#1} {\bf #2}, #3 (#4)}

\def\NIMA{{\em Nucl. Instrum. Methods} A}
\def\NPB{{\em Nucl. Phys.} B}
\def\PLB{{\em Phys. Lett.}  B}
\def\PRL{\em Phys. Rev. Lett.}
\def\PRD{{\em Phys. Rev.} D}
\def\ZPC{{\em Z. Phys.} C}
\def\EPC{{\em Eur. Phys. J.} C}

\def\be{\begin{equation}}
\def\ee{\end{equation}}
\def\bea{\begin{eqnarray}}
\def\eea{\end{eqnarray}}

\begin{document}
\vspace*{4cm}
\title{TRANSVERSE SPIN PHYSICS AT HERMES}

\author{A. Hillenbrand (on behalf of the HERMES Collaboration)}

\address{Physikalisches Institut II, Universit\"at
  Erlangen-N\"urnberg, Erwin-Rommel-Str. 1, 91058 Erlangen, Germany}

\maketitle\abstracts{
HERMES has measured azimuthal single-spin asymmetries of
charged pions produced in deep-inelastic scattering of positrons on
a transversely polarised hydrogen target.
The presented azimuthal moments provide access to two
yet unknown quark distribution functions, the transversity 
distribution function $\delta q$
and the Sivers function $f_{1T}^{\perp}$.
}

\section{Introduction}
\label{sec:intro}

In deep-inelastic scattering (DIS) the structure of the nucleon is probed
by the interaction of the target nucleon with a lepton beam. 
For a DIS process the four-momentum transfer $-Q^2$ (mediated at HERMES energies
of 27.6 GeV by the
exchange of a virtual photon) is large enough to resolve the nucleon's
constituents.
In the quark parton model, the virtual photons are assumed to 
scatter incoherently off the quarks in the nucleon. Thus the cross section can
be parameterised  by \emph{quark distribution functions}.
In leading twist only three distribution functions (DF) survive the integration
over the intrinsic transverse quark momentum $p_T$: the unpolarised
DF $q(x,Q^2)$ and the helicity DF $\Delta
q(x,Q^2)$ both of which have been investigated experimentally~\cite{martin,dq},
and the unknown transversity distribution $\delta
q(x,Q^2)$ ~\cite{t1,t2,t3,t4}. Here $x$ denotes the Bjorken scaling variable which can be 
interpreted as the fractional nucleon momentum carried by quark $q$. While the helicity distribution function
describes
the correlation of the quark spin with respect to the nucleon spin for
a longitudinally polarised nucleon, the transversity function $\delta
q$ gives the analogous correlation for a transversely polarised nucleon. 
These probabilistic
interpretations are valid in two different base systems: 
the helicity base ($\Delta q$), and the base of transverse spin
eigenstates ($\delta q$). These bases can be transferred into one another by rotation. 
Since nucleons are relativistic bound states, and in this regime boosts and rotations
do not commute, the helicity and transversity distributions may
differ.

In the helicity basis, transversity is related to a quark-nucleon
forward scattering amplitude involving helicity flips of both nucleon
and quark ($N^{\Rightarrow}q^{\leftarrow} \rightarrow
N^{\Leftarrow}q^{\rightarrow}$). Due to the chiral-odd nature of this
process, it can not be observed in inclusive measurements. However, in
semi-inclusive deep-inelastic scattering the transversity function can
be combined with another chiral-odd object, the Collins fragmentation
function (FF) $H_1^{\perp q}$ (\emph{Collins mechanism})~\cite{Collins}. 
This function describes the
correlation of the transverse polarisation of the struck quark with
the transverse momentum of the produced hadron. The combination of
$\delta q$ and $H_1^{\perp q}$ thus yields a left-right asymmetry (with
respect to the nucleon spin direction) in the
momentum distribution of the produced hadrons which can be observed as
azimuthal single-spin asymmetry (SSA). 

SSA can also be generated from the chiral-even Sivers function $f_{1T}^{\perp q}$. 
This function describes the correlation of the intrinsic transverse 
quark momentum $p_T$ with the transverse nucleon spin~\cite{Sivers}, and vanishes 
when integrating over the transverse quark momentum.
The Sivers function appears in the DIS cross section in combination 
with the spin-independent FF (\emph{Sivers mechanism}). 
In the helicity basis, the Sivers function is related to a scattering amplitude
representing a nucleon helicity flip without flipping the quark helicity 
($N^{\Rightarrow}q^{\leftarrow} \rightarrow N^{\Leftarrow}q^{\leftarrow}$) and thus 
must involve orbital angular momentum of the quarks~\cite{burkardt}.
SSA that can be attributed to $p_T$-dependent distribution functions such as the Sivers
function can also be understood in terms of final state interactions (FSIs) via a soft gluon~\cite{naive}.
These FSIs provide the mechanism to create the interference of amplitudes necessary to
establish the T-odd nature of the Sivers function.

\section{Experimental procedure}

\begin{figure}
	\begin{center}
		\psfig{figure=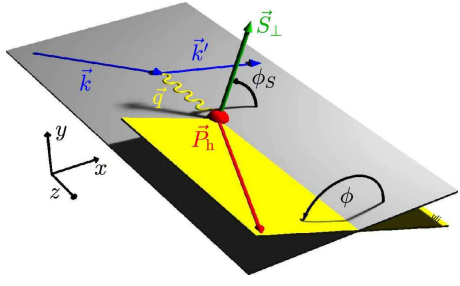,height=6cm,angle=270}
	\end{center}
	\caption{Definition of the azimuthal angles $\phi$ and $\phi_S$.}
	\label{fig:angles}
\end{figure}

The measurements presented here were performed using the HERMES detector~\cite{Hermes} located
at the HERA positron storage ring. Positrons with an energy of 27.6 GeV were scattered
off hydrogen gas injected into an open-ended tubular storage cell mounted
coaxial within the HERA beam line. The hydrogen atoms were transversely	polarised
with an average polarisation of $\langle P_z\rangle = 0.754 \pm 0.050$. The scattered positron
and the produced hadrons were detected with the HERMES spectrometer, which has a
momentum resolution of below 2.6\% and an angular resolution of below 1 mrad.
The average lepton identification efficiency is 98\%. A Ring-Imaging \v{C}erenkov detector
(RICH) provides the efficient separation of pions, kaons and protons over the momentum range
of 2 to 15 GeV, yielding a clean pion sample.

Both Collins and Sivers mechanisms lead to an azimuthal dependence of the production
of hadrons. Thus single-spin asymmetries have been extracted as a function of the two 
azimuthal angles $\phi$ and $\phi_S$ as
\bea
	A_{\mathrm{UT}}^{\pi^\pm} (\phi, \phi_S) &=& \frac{1}{\langle P_z\rangle}
		\frac{
			N_{\pi^\pm}^{\Uparrow}(\phi, \phi_S) - N_{\pi^\pm}^{\Downarrow}(\phi, \phi_S)
		}{
			N_{\pi^\pm}^{\Uparrow}(\phi, \phi_S) + N_{\pi^\pm}^{\Downarrow}(\phi, \phi_S)
		}\\
		&=& A_{\mathrm{UT}}^{\sin(\phi+\phi_S)} \sin(\phi+\phi_S) +
			A_{\mathrm{UT}}^{\sin(\phi-\phi_S)} \sin(\phi-\phi_S) + \dots
			\label{Eq:AUT}
\eea
The angles $\phi_S$ and $\phi$ describe the orientation of the target spin vector
and the hadron production plane relative to the lepton scattering plane, as shown in
Fig.~\ref{fig:angles}.
'UT' denotes unpolarised beam and transversely polarised target, $N_{\pi^\pm}^{\Uparrow(\Downarrow)}$ represents the luminosity-weighted yield of pions in
target spin state 'up' ('down'), and $\langle P_z\rangle$ is the average target polarisation. 
The amplitudes of the two different sine modulations~\cite{boer} in the SSA, caused by the Collins
($\sin(\phi+\phi_S)$) and the Sivers mechanisms ($\sin(\phi-\phi_S)$), were extracted 
simultaneously by a two-dimensional fit. This fit also took into account higher sine moments as
indicated by the dots in Eq.~\ref{Eq:AUT}.
The amplitudes contain the product of transversity distribution and Collins FF ($A_{\mathrm{UT}}^{\sin(\phi+\phi_S)}$) or of Sivers distribution and spin-independent FF ($A_{\mathrm{UT}}^{\sin(\phi-\phi_S)}$), respectively.

\begin{figure}
\begin{center}
  \psfig{figure=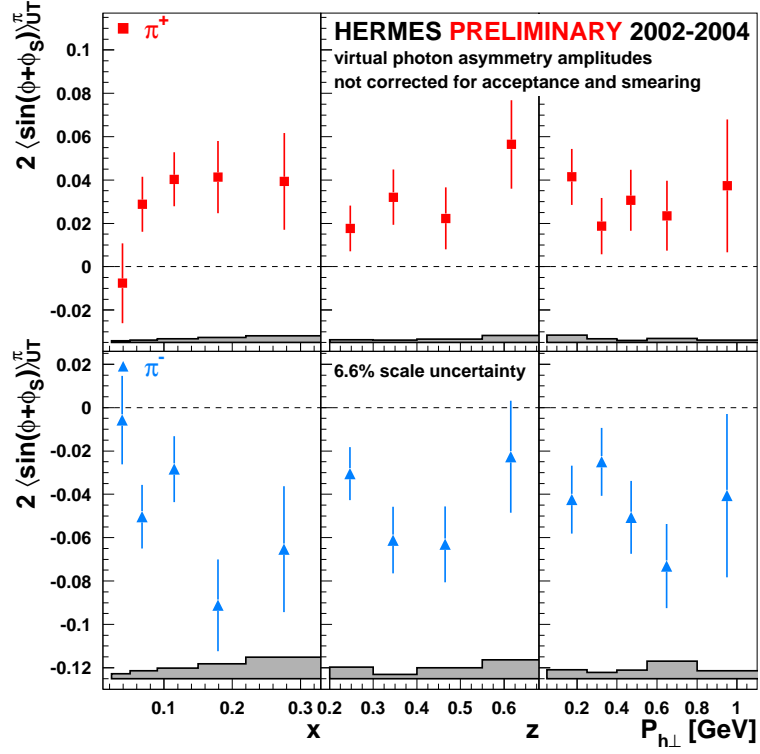,height=10cm}\\[0.5cm]
  \psfig{figure=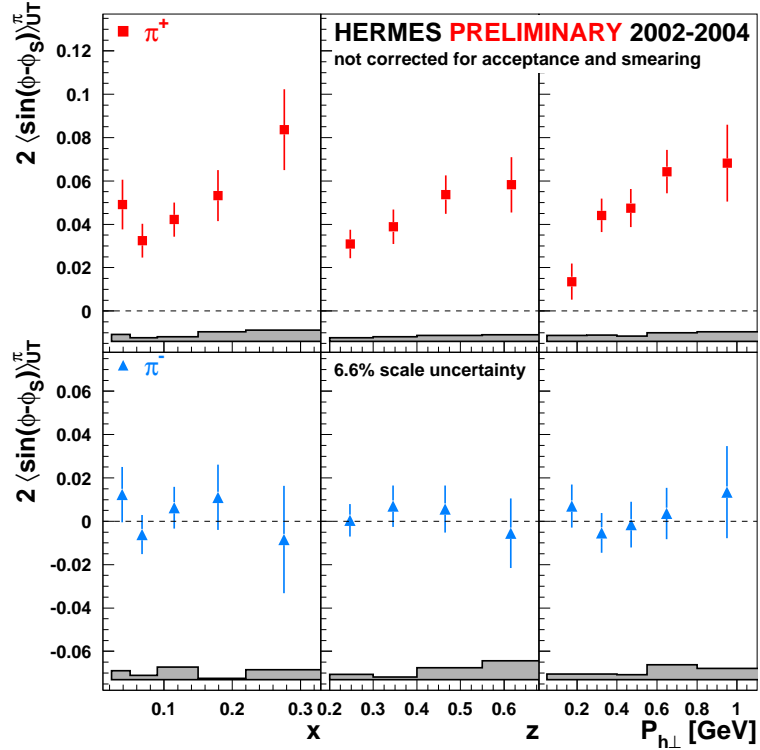,height=10cm}
\end{center}
\caption{Amplitudes for the Collins (upper panel) and Sivers (lower
  panel) mechanisms for positively and negatively charged pions as a function of $x$, $z$ and
  $P_{h\perp}$. The amplitudes are multiplied by a factor of two yielding
  an allowed range of $\pm1$. The error bands give the maximal
  systematic uncertainty due to acceptance and detector smearing
  effects as well as a possible contribution from the $\langle\cos
  \phi\rangle_{UU}$ moment in the unpolarised cross section. In
  addition, there is a general scale uncertainty of $6.6\%$ due to
  the uncertainty in the target polarisation.
\label{fig:CollinsSivers}}
\end{figure}

\section{Results}

Figure \ref{fig:CollinsSivers}  shows the amplitudes for the Collins
and Sivers moments for charged pions as a function of
$x$, $z$ and transverse hadron momentum $P_{h\perp}$. The Collins amplitude,
shown in the upper panel, is positive for $\pi^+$ and negative for $\pi^-$.
This is the first evidence for a non-vanishing Collins FF.
It is rather surprising that the $\pi^-$ amplitude is of at least the same magnitude as the $\pi^+$
amplitude,
suggesting a substantial disfavoured Collins function with opposite sign to that of
the favoured function.

The Sivers amplitude, shown in the lower panel, is significantly positive for $\pi^+$,
but consistent with zero for $\pi^-$. The positive $\pi^+$ moment is the first evidence
for the existence of a T-odd DF and is hence pointing to a non-vanishing orbital 
angular momentum of quarks.

\section{Outlook}

The $A_{\mathrm{UT}}^{\pi^\pm}(\phi, \phi_S)$ asymmetries shown here
represent combinations of distribution and fragmentation
functions. The ultimate goal is to extract the transversity and the Sivers DF. Parameterisations
of the spin-independent FF are available which will allow the
extraction of the Sivers function from the asymmetry moments.
Of particular interest is the determination
of the Sivers function from the Drell-Yan process, since the fundamental time-reversal properties of QCD should cause a sign change compared to DIS~\cite{drellyan}. 
It is expected that the transversity distribution
can soon be extracted using results on the Collins fragmentation function from BELLE~\cite{Ralf}. 

\section*{Acknowledgements}

This work has been supported by the German Bundesministerium f\"ur Bildung
und Forschung (BMBF) (contract nr. 06 ER 125I) and the European Community-Research
Infrastructure Activity under FP6 "Structuring of the European Research Area" program
(Hadron Physics I3, contract nr. RII3-CT-2004-506078).

\end{document}